\documentclass{elsart}

\usepackage{natbib,graphicx,amssymb}
\journal{New Astronomy}

\begin{document}

\begin{frontmatter}

\title{Modeling The Nucleosynthesis Of Massive Stars}

\author{T. Rauscher}

\address{Department of Physics and Astronomy, University of Basel,
Basel, Switzerland}

\begin{abstract}
This overview discusses issues relevant to modeling nucleosynthesis in
type II supernovae and implications of detailed studies of the ejecta. 
After a brief presentation of the most common approaches
to stellar evolution and parameterized explosions, 
the relevance of a number of nuclei to obtain information on the
evolution and explosion mechanisms is discussed. 
The paper is concluded by an outlook
on multi-dimensional simulations.
\end{abstract}


\end{frontmatter}

\section{Introduction}

Nowadays it is commonly accepted that stars with masses $M>8M_\odot$
complete all possible phases of hydrostatic burning up to Si burning
and end their life in a collapse of the Fe core, followed by an explosive
ejection of matter. The most common type of core collapse supernovae
are type II supernovae, showing H lines in their spectrum and resulting
from the explosion of a progenitor with $M>10M_\odot$.
The details of hydrostatic stellar evolution depend
on hydrodynamical effects like convection as well as on nuclear physics.
Regarding the latter, the most famous example is the one of the
$^{12}$C($\alpha$,$\gamma$)$^{16}$O reaction which sensitively
determines energy generation during and the $^{12}$C/$^{16}$O ratio
after He burning. Therefore, also the subsequent burning phases depend
sensitively on this reaction and thus the evolution of the star
\citep[see, e.g.,][]{boy02}. 
Due to the still large uncertainty in the cross
section, this reaction contributes the largest uncertainty in
hydrostatic stellar evolution and its nucleosynthesis. More recently, it
has been shown that two other reactions are also of major importance,
namely the ones of the branching between $^{22}$Ne($\alpha$,n)$^{25}$Mg
and $^{22}$Ne($\alpha$,$\gamma$)$^{26}$Mg. The neutrons released by the
former reaction alter abundances by neutron capture reactions and are
responsible for the weak s-process component produced in massive stars.
At sufficiently high temperatures the neutron release can be so
effective that nuclei several units away from stability can be created
in a so-called n-process \citep{rau02}. On the other hand,
prevailing uncertainties in the treatment of convection and
semi-convection will influence the reaction of the star to altered
nuclear physics. While nuclear reactions dominate the error in
calculating hydrostatic abundances, the proper hydrodynamic treatment of
the core collapse currently seems to be the source of most problems
connected with the explosion mechanism. Nevertheless, nuclear properties
like neutrino cross sections and the nuclear equation of state have to
be known accurately, too.

Observing the abundances of radioactive species in supernova remnants
help us gain a deeper understanding concerning the production of
those nuclei and also on the underlying physical processes. Looking at
nuclei produced in different phases of evolution and explosion will
probe different aspects. Therefore it is important to distinguish the
relevant nucleosynthetic processes.

\section{How to model nucleosynthesis}

In principle, multi-dimensional (multi-D)
hydrodynamical calculations are necessary to
follow convection, mixing and especially the explosion. However, due to
limitations in both computer power and numerical approaches, it is not
yet possible to couple full reaction networks, including all nuclei ever
produced in such stars, to multi-D hydrodynamical solvers. Moreover,
even multi-D models have problems describing the explosion mechanism
because they currently do not show explosions at all 
\citep[see, e.g.,][]{bur03}. Focussing on nucleosynthesis, one
traditionally resorts to a number of approximations: Instead of multi-D,
the simulation is reduced to one spatial dimension; two reaction
networks are used, a smaller one which provides the nuclear energy
generation and is directly coupled to the hydro solver, and a larger one
without feed-back to hydro but carrying all the nucleosynthesis; multi-D
effects such as convection and mixing are treated in approximations,
like mixing length theory, and by invoking convection criteria
(Schwarzschild, Ledoux); the explosion itself is parameterized. Even
which such approximations it became only recently possible to
consistently follow synthesis of all nuclei up to Bi in a single, large
network \citep{rau02}. Despite the limitation, one expects mostly
reliable results for nucleosynthesis of nuclei independent of the
explosion mechanism, with the exceptions mentioned in the Introduction.
This has been nicely proven by comparison with observational data, both
regarding yields and velocity distributions of the ejecta.

There are mainly two ways to parameterize the explosion. The first is to
artificially increase the entropy in the core of an evolved progenitor.
This is done by a sudden temperature enhancement in the Fe core,
resulting in an increase in pressure, a shock wave, and finally ejection
of the outer layers. The mass cut, i.e. the mass coordinate separating material
which will fall back from the one really ejected, is another free
parameter in this type of simulation. The induced thermal energy is
adjusted as to reproduce observed explosion energies whereas the mass
cut can be chosen to be in accordance with abundances of nuclei produced
in the explosion in the inner-most layers, such as $^{56}$Ni. 
This approach is applied by, e.g., \citet{TNH96} and co-workers.

Another way to create an artificial explosion is to input kinetic energy
by a one-dimensional moving piston. The piston first moves inward with a
fraction of the local gravitational acceleration as the
core collapses and then outward in another ballistic trajectory
to induce a shock. The outward acceleration is usually considered the
only open parameter in this model. The mass cut is implicitly obtained 
by the mass settled on the piston after a sufficiently large time.
The free parameter thus can either be determined by reproducing the
observed kinetic energy in the ejecta or by the ejected
amount of inner material, such as $^{56}$Ni. The latter is usually used
for stars with $M>20M_\odot$ because those usually have higher Ni
yields. It should be noted,
however, that there are two further implicit parameters, the inward
piston acceleration and the initial position of the piston, which are
usually treated as fixed. The fall back, and thus the $^{44}$Ti and
$^{56}$Ni yield, is not only dependent on the piston energy but also
on the latter. This approach was introduced by \citet{WW95} and
subsequently used by that group 
\citep[for recent examples, see][]{hoff01,rau02}. Very recently, 
also \citet{lim03} adopted it.

Obviously, both approaches do not account for the detailed collapse
process and therefore cannot provide a consistent description of the
expected neutrino pulse. In order to study neutrino-induced
nucleosynthesis, usually parameterized neutrino burst profiles are
applied, mostly influencing light element nucleosynthesis of Li, B, F,
and partially also of $^{138}$La and $^{180}$Ta in the $\nu$-process
\citep{heg03}. Both approaches
also cannot follow the innermost high-entropy convective zones thought
to be a possible site of the r-process. R-process abundances cannot
be obtained although the thermal approach yields a slightly better
description of the entropy in the lowest shells. It should be noted,
however, that also more self-consistent multi-D simulations are not able
to obtain the entropies required for the r-process. Whether this is an
indication that supernovae are not the site of the r-process or whether
this reflects deficiencies in the modeling, perhaps related to the
failing explosions, remains an open question. For further considerations
concerning problems with the r-process in type II supernovae, see
\citet{frei}.

For completeness, a third approach has to be mentioned which is, to my
knowledge, only rarely used: the radiation dominated shock
approximation \citep{ww80,arn96}. It is a simpler description of 
the outgoing shockwave than the above approaches. Until recently, it was
used by \citet{cls,lsc} and co-workers. See \citet{lim03} for a
comparison to the piston approach.

\section{Nuclide classes}

Different nuclides probe different aspects of stellar evolution and
explosion. One can define three coarse classes. In the following,
examples for nuclei in each class are provided but that is by no means meant
to be a complete list. The yields of species
in the first class are determined by stellar evolution only, they are
mainly produced in hydrostatic burning but their abundances can also be
altered by explosive burning in the supernova shock front. 
They are sensitive to uncertainties
in the reaction rates and to mixing effects as given by the stellar
structure. Their yields vary with the mass of the progenitor star. Such
elements are He, C, O, Ne, Mg. Among the radioactive species are
$^{26}$Al, $^{59}$Co, $^{60}$Fe. It is interesting to note that there is
no experimental determination of the rate of the
reaction $^{59}$Fe(n,$\gamma$)$^{60}$Fe, producing $^{60}$Fe. Therefore,
its yield also has a considerable nuclear uncertainty.

The second class comprises nuclei whose yields depend on stellar
evolution as well as the explosion energy. They are only weakly
dependent on the progenitor mass. Examples are isotopes of Si, S, Ar,
Ca.

The yields of the nuclear species in the final class probe the explosion
mechanism. They depend on the size of the pre-supernova Fe core, the
assumed mass cut, the explosion energy, and on the electron abundance
$Y_e$ which provides a measure of the neutronization of the matter. The
nuclei in this class are those from $^{44}$Ti to mostly Fe-group nuclei
(including $^{56,57}$Ni). Also r-process nuclei would fall in this
category but they cannot be treated in the parameterized models
introduced above. This nuclide class can be used to fix the model
explosion parameters. However, $Y_e$ in the inner zones can be altered
by neutrino-induced weak interactions. As mentioned before, this effect
is not included and therefore the obtained parameters are rather
effective parameters than actual measures of physical quantities. In
this context it is arguable whether a higher number of free parameters
is a lack of consistency or a merit, providing more flexibility.

\section{Prospects}

Future improvements in parameterized models will concern the size of the
reaction networks 
\citep[specifically also the one used for energy generation,][]{woo03}
and the treatment of convection. Better constrained nuclear reaction
rates would also provide a major improvement. Improved parameterizations
of the explosion and the neutrino pulse within the discussed limitations
can be obtained from comparisons with observation and multi-D models.
First steps have been taken \citep[e.g.][and Travaglio et al., this
volume]{kif99} to couple nucleosynthesis networks to 2-D simulations. As
in early 1-D simulations one has to resort to very limited networks yet,
and it is not clear whether one has to go to higher dimensions to
properly model the convective flows. Similar
to the 1-D approaches, an artificial explosion has to be invoked since
self-consistent calculations still offer little guidance as to the exact
placement of the mass cut, the entropy and $Y_e$ of the innermost
ejecta, or even if a given model will explode \citep{her94,jam96,fry00}.
Nevertheless, multi-D effects such as mixing and asphericity can be
studied in such models. These can have two consequences. If the explosive
nuclear burning zones (i.e.\ the shock wave) become non-spherical,
explosive nucleosynthesis would be altered, leading to different
explosive yields. However, recent calculations still show a mainly
spherical shock propagation \citep{kif99}. The second consequence
concerns the mixing behind the burning front. It will not directly
affect nucleosynthesis but the burning products will be mixed into
different layers behind the shock front, 
affecting the observational signature (see Travaglio
et al., this volume).

It should be kept in mind that any exhaustive
investigation of the origin of the elements has to consider, among
others, a variety of progenitor stars with different initial masses and
metallicities. Despite of the progresses in multi-D simulations,
nucleosynthesis studies with full reaction networks on an extensive
grid of masses
and metallicities are only feasible with parameterized 1-D models, yet.
Therefore, such models will stay with us for a while.

{\bf Acknowledgement:}
TR is supported by the Swiss NSF
with a PROFIL professorship (grant 2024-067428.01) and through a
research grant (2000-061031.02).


\end{document}